# Anisotropic skyrmion diffusion controlled by field-induced symmetry breaking

Nico Kerber[1,2], Markus Weißenhofer[3], Klaus Raab[1], Kai Litzius[1,2,4], Jakub Zázvorka[1,5], Ulrich Nowak[3], Mathias Kläui[1,2]†

[1]*Institut für Physik, Johannes Gutenberg-Universität Mainz, 55099, Mainz, Germany.*
[2]*Graduate School of Excellence Materials Science in Mainz, 55128 Mainz, Germany.*
[3]*Fachbereich Physik, Universität Konstanz, Universitätsstraße 10, 78457 Konstanz, Germany*
[4]*Department of Materials Science and Engineering, Massachusetts Institute of Technology, Cambridge, Massachusetts 02139, USA*
[5]*Institute of Physics, Faculty of Mathematics and Physics, Charles University, Ke Karlovu 5, 121 16, Czech Republic.*

†Corresponding author: klaeui@uni-mainz.de

**Abstract**

Diffusion of particles has wide repercussions ranging from particle-based soft matter systems to solid state systems with particular electronic properties. Recently, in the field of magnetism, diffusion of magnetic skyrmions, topologically stabilized quasi-particles, has been demonstrated. Here we show that by applying a magnetic in-plane field and therefore breaking the symmetry of the system, the skyrmion diffusion becomes anisotropic with faster diffusion parallel to the field axis and slower diffusion perpendicular to it. We furthermore show that the absolute value of the applied field controls the absolute values of the diffusion coefficients so that one can thereby uniquely tune both the orientation of the diffusion and its strength. Based on the stochastic Thiele equation, we can explain the observed anisotropic diffusion as a result of the elliptical deformation of the skyrmions by the application of the in-plane field.

Magnetic skyrmions, topologically stabilized spin structures, allow for a wide range of promising room temperature applications in logic, data storage, probabilistic computing, and many other areas [1-6]. Skyrmions can be stabilized by the chiral Dzyaloshinskii-Moriya Interaction (DMI) [7-8] arising in bulk systems with broken inversion symmetry such as B20 compounds [9-11], as well as in thin multilayered films with structural inversion asymmetry [12-14]. It has been demonstrated that it is possible to reproducibly write and delete single skyrmions [14-15]. Furthermore, reproducible skyrmion displacement in heavy-metal/ferromagnet thin film heterostructures by the highly efficient spin orbit torques (SOT) generated through the interface that also provides the homochiral character of the spin textures, has been thoroughly investigated [15-19]. While current-induced skyrmion dynamics has been a major field of study in skyrmionics research due to the promising applications in storage and logic devices, studies investigating thermally induced skyrmion dynamics have been elusive so far. This has been frequently based on the common view that thermal diffusion is strongly suppressed in magnetic systems in which the damping is typically small [20]. It has just recently been shown that in specially tailored low-pinning multilayers pure isotropic skyrmion diffusion at room temperature can be achieved [6, 21-23]. By changing the sample temperature, one obtains control over the diffusion strength, since the diffusion coefficient depends exponentially on the temperature in case of a non-flat energy landscape as present in the experimental samples [6]. However, in order to use skyrmion diffusion, one needs to control and for instance confine the diffusion along a certain direction which can then possibly enable an implementation of skyrmion based Brownian circuits [24].

While diffusion of skyrmions is a recent finding, it is well established in the field of soft matter, where diffusive motion of particles [25] has been observed already in 1827 by Robert Brown and was first theoretically described by Albert Einstein, who established a relation between the friction coefficient of a Brownian particle and its diffusion coefficient [26]. In the extensive subsequent investigations, it was shown that uniaxial fabricated ellipsoid particles exhibit stronger diffusion parallel than perpendicular to their long axis leading to anisotropic translational Brownian diffusion [27-28]. This suggests that one can control diffusion by changing the particle shape. Nevertheless, in soft matter systems one cannot change the particle shape after the experiment has started so that there is limited control during the experiment. However, in magnetic systems, a flexible dynamic change of the skyrmion shape could be envisaged as a unique approach that provides flexible dynamic control of skyrmion diffusion.

Thus, in this study, we explore the effects of applied fields as means to control the diffusive motion of skyrmions in a CoFeB-based thin film metal-multilayer system with perpendicular magnetic anisotropy. By Kerr microscopy we determine the diffusion properties under the application of an in-plane (IP) field oriented along a selected direction. We find that application of in-plane fields leads to anisotropic diffusion allowing one to control the diffusion direction and amplitude over a wide range, which can be explained theoretically based on an extended Thiele equation by a combination of analytical and numerical calculations.

The investigated Ta(5 nm)/Co$_{20}$Fe$_{60}$B$_{20}$(1 nm)/Ta(0.08 nm)/MgO(2 nm)/Ta(5 nm) materials stack was deposited via dc-magnetron sputtering on a Si/SiO$_2$ substrate and the growth details as well as the material characterization can be found in Ref. [6] where skyrmion diffusion was established in nominally identical samples. We obtain a typical hour-glass shaped hysteresis loop indicating the presence of a magnetic stripe or skyrmion phase [29]. A low effective anisotropy and saturation magnetization allow substantial changes in the domain alignment by small magnetic field changes.

By changing the applied out-of-plane (OOP) field we can tune the skyrmion diameter, which we set to about 1.5 µm by fixing the OOP-field to 0.3 mT. The topologically non-trivial character of the skyrmions has been checked by displacement via spin-orbit torques [16]. In order to image the diffusion dynamics, we use a magneto-optical Kerr effect (MOKE) microscope with a time resolution of 62.5 ms. To achieve significant diffusion that is still detectable within the given time resolution we set the sample temperature to 303 K using a Peltier element, leading to an increased diffusion coefficient compared to room temperature [6]. We use ImageJ's TrackMate plugin [30] to identify and track skyrmions during their diffusive motion. The hereby obtained skyrmion tracks are then used to investigate the time dependence of the mean-squared displacement that is fitted using a linear dependence to obtain the diffusion coefficients [6].

To go beyond the previous observation of isotropic diffusion [6], we need to break the in-plane symmetry. To this end, we probe the influence of an in-plane field along a selected direction on the diffusion properties. To visualize the effect, we assume the applied field direction as the x-axis and we now separately show in Fig. 1 the projected mean squared displacement (MSD):

$$\mathrm{MSD}_x = \left\langle \left(x(t) - x(0)\right)^2 \right\rangle = 2\mathrm{D}_x t \qquad (1)$$

parallel (MSD$_x$) and perpendicular (MSD$_y$) to the x-axis IP-field as a function of time.
Firstly, we observe that at zero IP-field the MSD parallel and perpendicular to the IP-field axis overlap within the error bars, as expected for the case of isotropic diffusion.
Next, we study the effect of the IP-field systematically by measuring the diffusive motion for several IP-field values. We calculate the diffusion coefficient parallel (D$_x$) and perpendicular (D$_y$) to the IP-field axis according to Eq. (1) as shown in Fig. 2.
The first striking observation is that upon IP-field application, a significant decrease of the absolute value of the diffusion coefficient occurs (Fig. 2(a)). Therefore, as a first key result, we can tune the magnitude of the diffusion coefficient over a sizeable range dynamically, while previous handles to change the diffusion coefficient like temperature [6] are less practical as a fast and flexible control parameter.
As the second key result, we see strongly different diffusive behavior parallel and perpendicular to the IP-field axis (Fig. 2(a)). The ratio of the diffusion coefficients D$_x$/D$_y$ is shown in Fig. 2(b). Starting at a ratio of 1 indicating isotropic diffusion at zero IP-field, we find that above a threshold field of about 1 mT, the ratio starts to increase up to a value of (1.46±0.27) at the maximum IP-field of 2 mT used in this study. For even higher IP-fields the skyrmions in these samples are destabilized by the IP-field and start to break up into stripes so that we cannot reliably determine the skyrmion diffusion.

We therefore find that by applying an in-plane field in a selected direction, we can generate anisotropic diffusion. This provides a new means to steer the thermal skyrmion motion. This is interesting not just from a fundamental science perspective but also for potential applications, where in previous studies the authors have patterned geometrically a skyrmion circuit to steer diffusive skyrmion motion [22, 24].

As the next step we need to understand the origin of the observed anisotropic diffusion dynamics. Fundamentally, there are different possible origins as the in-plane field changes the skyrmion shape, the domain wall delineating the skyrmion, as well as the average in-plane magnetization in the sample. In the field of soft matter, it is known that particles with an anisotropic shape such as elongation show anisotropic diffusion with different diffusion coefficients along the long and the short axis [27-28]. Drawing on this experience, we study the shape of the skyrmion and determine the skyrmion diameter size parallel and perpendicular to the IP-field axis. We fit each tracked skyrmion in each frame by a 2D Gaussian function:

$$f(x,y) = A\, exp\left(-\left(\left(\frac{(x-x_0)^2}{2\,\sigma_x^2}\right) + \left(\frac{(y-y_0)^2}{2\,\sigma_y^2}\right)\right)\right) \quad (2)$$

with the size $s_x = 2\sqrt{2\,ln(2)}\,\sigma_x$ parallel and $s_y = 2\sqrt{2\,ln(2)}\,\sigma_y$ perpendicular to the IP-field axis given by the FWHM. We note that due to the diffusive motion, as well as the influence of pinning sites the skyrmion shape might be distorted in single frames, while the small error bars presented in Fig. 3 are based on the statistics from averaging over all skyrmions and frames. Averaging over all frames leads to the field dependence of the skyrmion size shown in Fig. 3. While the size perpendicular to the field direction just slightly decreases with increasing IP-field, one can observe a strong increase of the skyrmion size above a field of ca. 1 mT as shown in Fig. 3(a). This in turn leads to an increase of the size ratio above a threshold IP-field of ca. 1 mT as shown in Fig 3(b). We note that the threshold field for the skyrmion size is in line with the observed threshold field for the diffusion coefficients.

As a next step we have to thus check theoretically if our result of an anisotropic shape is expected to lead to anisotropic diffusion.

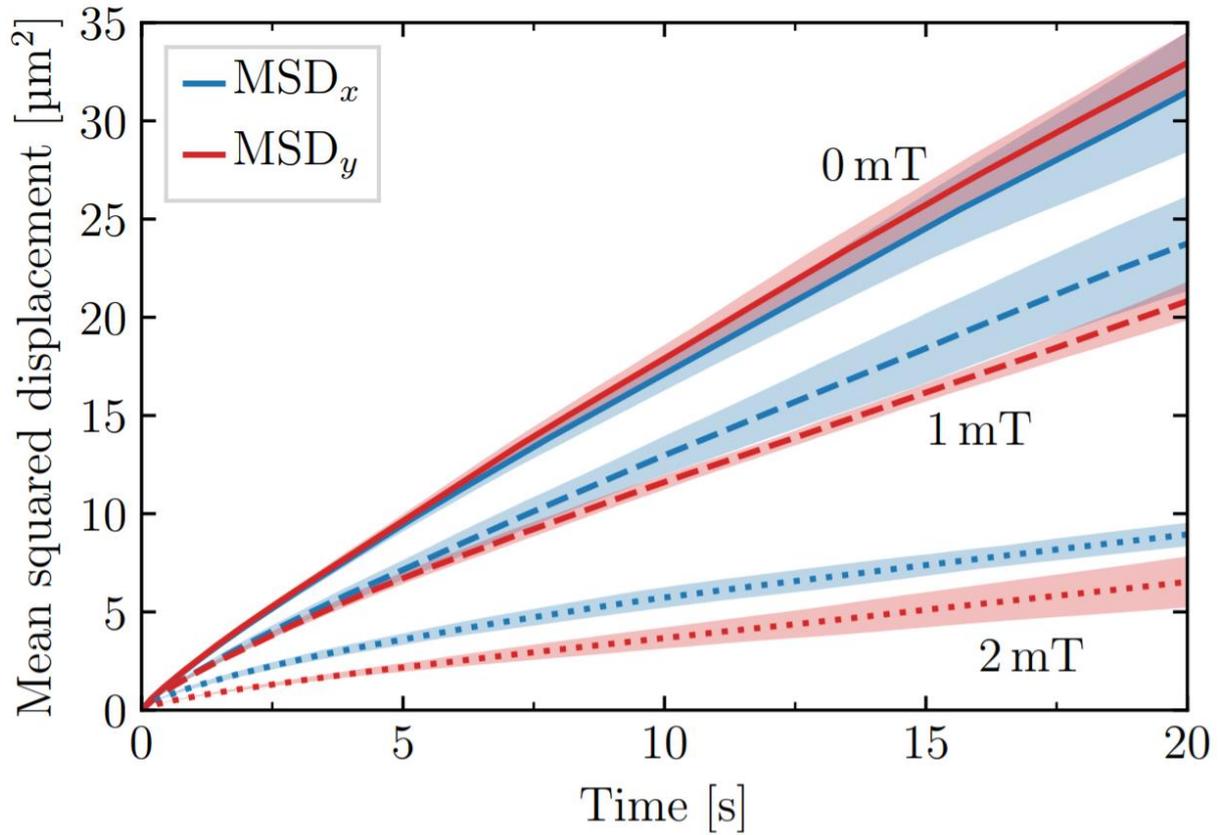

**Figure 1.** Mean squared displacement MSD$_x$ (blue) and MSD$_y$ (red) as a function of time under application of a magnetic in-plane field of 0 mT (solid), 1 mT (dashed) and 2 mT (dotted). While the displacement without IP-field application shows isotropic diffusion within the error bars, application of an IP-field leads to an increasing anisotropic behaviour.

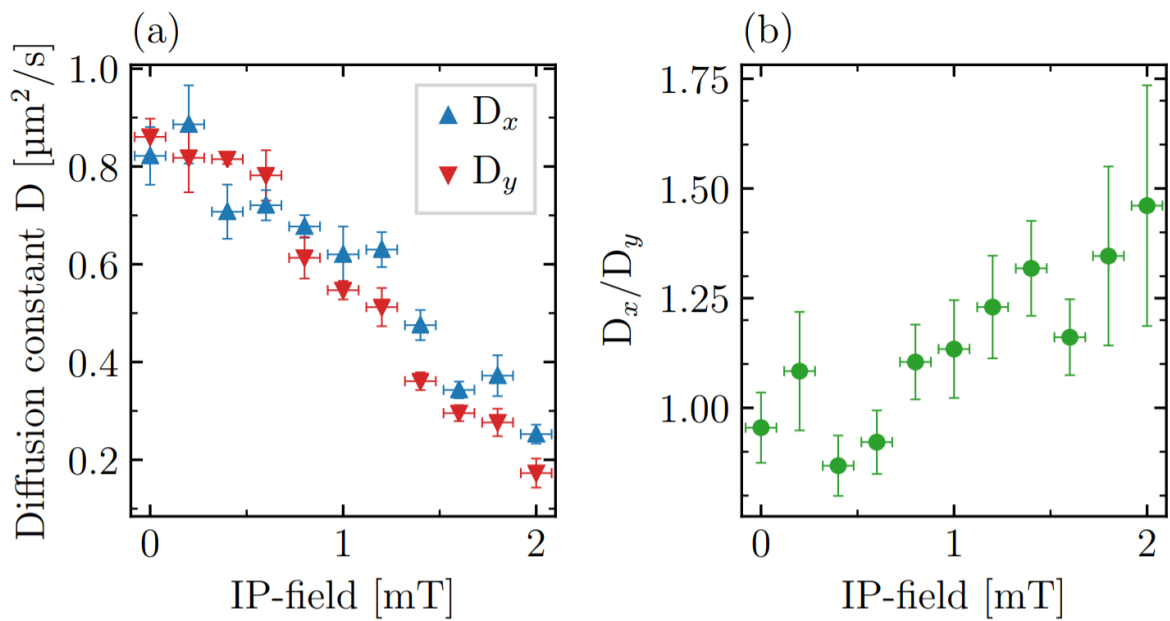

**Figure 2.** (a) Diffusion coefficients $D_x$ and $D_y$ as a function of magnetic in-plane field. The application of an IP-field leads to a decrease of the diffusion coefficients as well as (b) an increase of the ratio of the diffusion coefficients $D_x/D_y$ identifying anisotropic diffusion.

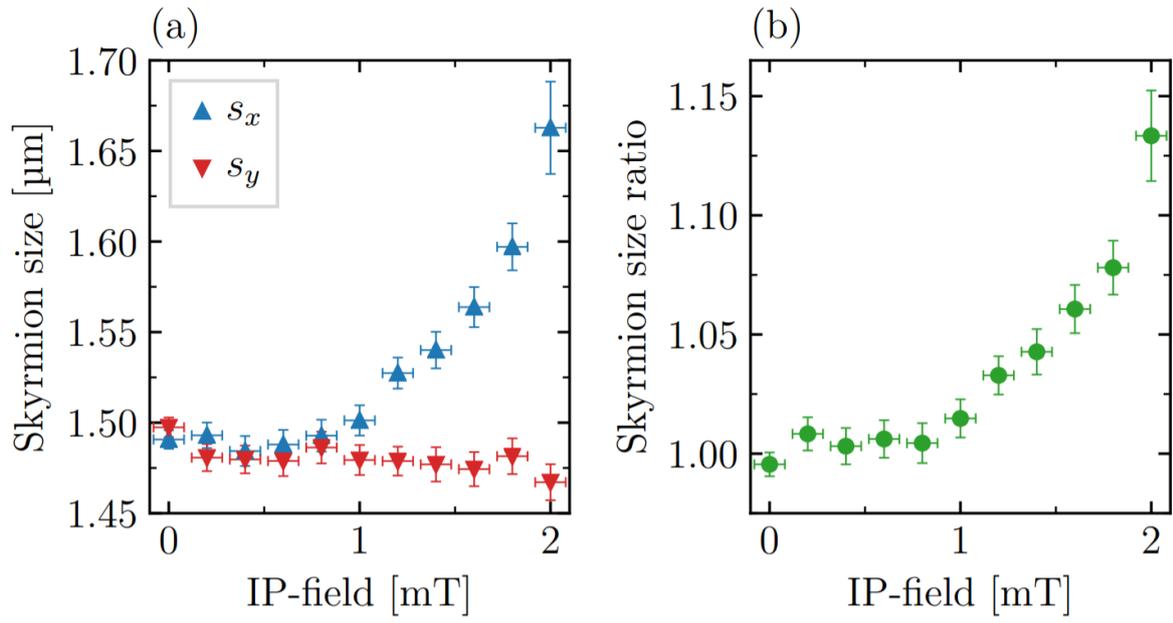

**Figure 3.** (a) Skyrmion size $s_x$ and $s_y$ as a function of magnetic in-plane field. The application of an IP-field leads to an elongation of the skyrmion parallel to the field axis. (b) This subsequently leads to an increase of the size ratio $s_x/s_y$ upon IP-field application.

To investigate theoretically the influence of the skyrmion shape on the diffusive dynamics of skyrmions we start with an analytical model of skyrmion diffusion. In the first theoretical study [20] of thermal skyrmion diffusion the following expression for the diffusion coefficient of a rotationally symmetric skyrmion in a rotationally and translationally invariant two-dimensional film was derived from Thiele's equation [31]:

$$D^{\text{iso}} = k_B T \frac{\alpha D_T}{(\alpha D_T)^2 + G^2} \tag{3}$$

with the gyrocoupling $G = M_{2d}/\gamma \int \mathbf{S} \cdot (\partial_x \mathbf{S} \times \partial_y \mathbf{S}) d^2 r = 4Q\, M_{2d}/\gamma$ and $2D_T$ being the trace over the dissipative tensor $D_{ij} = M_{2d}/\gamma \int \partial_i \mathbf{S} \partial_j \mathbf{S} d^2 r$. Note, that only the dissipative tensor depends on the skyrmion profile whereas the gyrocoupling is a topological invariant. Eq. (3) describes isotropic diffusive motion with $\text{MSD}_x = \text{MSD}_y = 2\, D^{\text{iso}} t$ and its validity was demonstrated via atomistic spin simulations [6, 32].

However, if the rotational symmetry of the magnetic texture is broken, the dissipation tensor becomes anisotropic which consequently leads to different dynamics along different axes of the skyrmion. For current driven motion of spin structures with broken rotational symmetry the anisotropy of the skyrmion dynamics was recently predicted [33-34].

As anisotropic diffusion dynamics has so far not been treated in literature, we here derive the generalization of Eq. (3) for a general dissipative tensor starting from Thiele's equation. Thiele's equation describes the movement of a rigid magnetic texture and reads as $G \times v + \alpha D v = F$ where $G \times v$ is the gyrotropic force, $\alpha D v$ is the dissipative force and $F$ is an applied force. In the case of thermal diffusion, the applied force is a fluctuating, stochastic force leading to a Brownian-type motion of the skyrmion. In the so-called "low" noise limit its autocorrelation is given via [35]:

$$\langle F_i(t) F_j(t') \rangle = 2\alpha k_B T D_{ij} \delta(t - t'). \tag{4}$$

By expressing the cross product in the gyrotropic force via a skew-symmetric matrix skew[$G$] Thiele's equation can be rewritten as:

$$v = (\text{skew}[G] + \alpha D)^{-1} F \tag{5}$$

which, together with Eq. (4), allows for the calculation of the velocity autocorrelation function $\langle v_i(t) v_j(t') \rangle$. Finally, time integration $r(t) = \int_0^t v(\tau) d\tau$ yields for the diffusion tensor

$$D = \begin{pmatrix} \langle x^2(t) \rangle & \langle xy(t) \rangle \\ \langle xy(t) \rangle & \langle y^2(t) \rangle \end{pmatrix} / 2t = \frac{k_B T}{\alpha^2(D_{xx} D_{yy} - D_{xy}^2) + G^2} \begin{pmatrix} \alpha D_{yy} & -\alpha D_{xy} \\ -\alpha D_{xy} & \alpha D_{xx} \end{pmatrix}. \tag{6}$$

As the diffusion tensor and the dissipation tensor are both diagonalizable and connected via Eq. (6), we will choose for the following discussion a coordinate system for which the off-diagonal terms in both tensors vanish, i.e. $\langle xy(t) \rangle = 0$. For simplicity, the diagonal elements of the diffusion tensor will be referred to as $D_x$ and $D_y$ in what follows, with $\langle x^2(t) \rangle = 2D_x t$ and $\langle y^2(t) \rangle = 2D_y t$.

The anisotropy of the diffusive motion can most readily be characterized by the ratio $D_x/D_y$. Eq. (6) predicts that this ratio can be expressed as $D_{yy}/D_{xx}$ which implies anisotropic diffusion for spin textures with $D_{yy} \neq D_{xx}$. In most systems, skyrmions are found to be of circular shape resulting in isotropic behavior with $D_{xx} = D_{yy}$. However, it has been recently demonstrated [33, 36] that an applied in-plane field can lead to an anisotropic skyrmion profile. When we now feed this into our generalized calculation, this consequently leads to different dissipation tensor elements for the directions along the in-plane field and orthogonal to it.

As the deformation can be continuously tuned by varying the strength of the applied in-plane field, this should, according to theoretical predictions derived above, allow for a transition from isotropic to anisotropic diffusive motion. While we can thus qualitatively understand the experimental results, we next test the analytically derived dependence of the skyrmion diffusion on the skyrmion profile by micromagnetic simulations using Mumax [37]. The following magnetic material parameters were used: Dzyaloshinskii-Moriya interaction (D = 0.0985 mJ/m$^2$), exchange stiffness constant (A = 2.46 10$^{-12}$ J/m ), saturation magnetization (M$_S$ = 9.8 10$^5$ A/m), perpendicular magnetic anisotropy (K$_{eff}$ = 6.4 10$^4$ J/m$^3$). The skyrmion profile was equilibrated under the application of different IP-fields at an OOP-field of 0.2mT. The size parallel (x-axis) and perpendicular (y-axis) to the IP-field axis is shown in Fig. 4. The skyrmion elongates strongly parallel and shortens slightly perpendicular to the IP-field above a certain threshold field of ca. 0.75 mT. We here find not only qualitative agreement but also quantitative agreement with the experimentally observed behavior, when the used parameters are also comparable to the ones obtained experimentally. The size ratio s$_x$/s$_y$ shown in Fig. 4(b) reproduces the experimental results with a flat behavior up to the threshold of about 0.75 mT followed by an increase of this ratio.

Next the simulated skyrmion profile was used to numerically calculate the dissipation tensor elements $D_{xx}$ and $D_{yy}$ to obtain also quantitative insights into the expected anisotropy of the diffusion. Since the dissipation tensor elements are directly connected to the diffusion coefficients D$_x$ and D$_y$ as derived above we directly calculate these as presented in Fig. 5(a), where we assume that α = 0.01. Based on the obtained size-field dependence, we obtain above the threshold field of 0.75 mT a slightly increased diffusion parallel to the IP-field axis while the diffusion coefficients decrease strongly perpendicular to the IP-field. We find that the numerically obtained ratio D$_x$/D$_y$ of the diffusion coefficients shown in Fig. 5(b) agrees with the experimental results.

Thus, by comparing theory and experiment, we identify that the origin of the observed anisotropic diffusion induced by an IP-field is the deformation of the skyrmion shape due to this field. Analytical calculations reproduce the trend that non-spherical skyrmions lead to anisotropic diffusion due to non-symmetric dissipation tensor elements and numerical calculations allow us to quantitatively reproduce the experimentally found skyrmion shape deformation, which when entered into the analytical calculation reproduces the anisotropy of the diffusion even quantitatively with good fidelity.

Finally, we note that our theory does not reproduce all details of the experimental results. In particular, we are not able to reproduce the observed strong decreases of both diffusion coefficients upon IP-field application. This means that there needs to be an additional mechanism, which scales both diffusion coefficients in a similar fashion leading to a strong decrease of both coefficients upon IP-field application while preserving the anisotropic diffusion ratio. A possible source of this discrepancy is the fact that theoretically, we have considered free skyrmion diffusion in a flat energy potential landscape. In a non-flat energy potential landscape, it has been shown that the influence of pinning sites leads to a clear exponential dependence of the diffusion coefficient on temperature and by that to a deviation from the pinning free case [6]. As pinning of the skyrmions can be affected by the change in the shape and size, this could then affect the absolute values of the diffusion coefficients. The details of this are however beyond the scope of the current investigation and could be studied in future work.

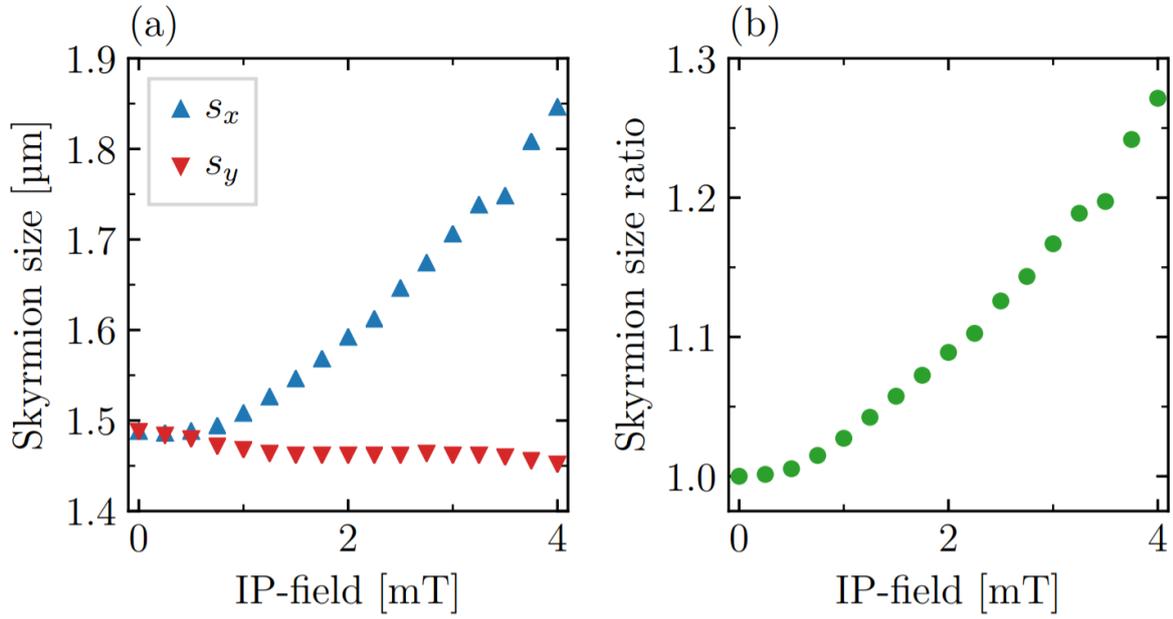

**Figure 4.** (a) Numerically simulated skyrmion size $s_x$ and $s_y$ as a function of magnetic in-plane field. The application of an IP-field leads to an elongation of the skyrmion parallel to the field axis and a decrease perpendicular to it. (b) This in turn increases the size ratio $s_x/s_y$ upon IP-field application.

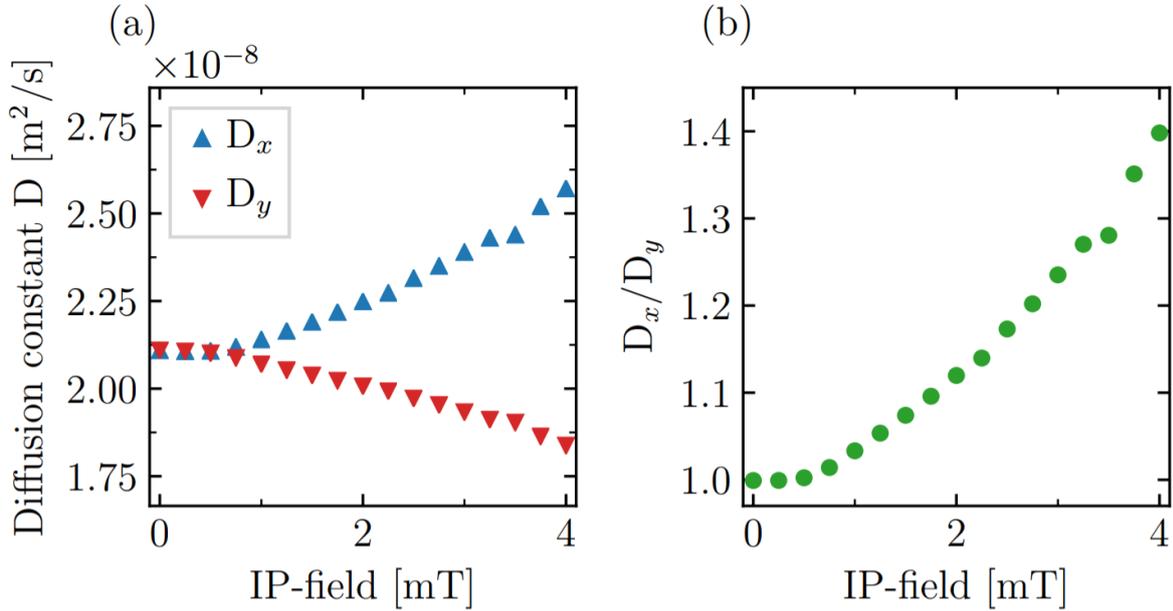

**Figure 5.** (a) Calculated diffusion coefficients $D_x$ and $D_y$ following Eq. 6 using the numerically determined dissipative tensor elements as a function of magnetic in-plane field. The application of an IP-field leads to an increase of the diffusion coefficient parallel to the IP-field axis and a decrease perpendicular to it. (b) An increase of the ratio of the diffusion coefficients $D_x/D_y$ and therefore anisotropic diffusion upon IP-field application follows.

In conclusion, using time resolved MOKE microscopy we observe skyrmion diffusion whose magnitude and anisotropy we control by the strength and direction of an applied IP-field. We identify a dependence of the anisotropy of the skyrmion shape on the IP-field that is similar to the field-dependence of the diffusion anisotropy. To understand this, we perform analytical calculations, from which we obtain a connection between the diffusion tensor and the dissipation tensor, which depends on the skyrmion shape. Carrying out additional micromagnetic simulations and numerical calculations we obtain qualitative as well as quantitative agreement with the experimentally observed behavior. Thus, our findings reveal how by tuning the shape of the skyrmion by a magnetic field we are able to control the diffusive dynamics. Given the fact that the skyrmion shape can be controlled by a variety of additional means including dynamic strain and strongly elliptical skyrmions have been observed [38], our findings open a range of possible handles to control diffusive dynamics. And it is exactly this type of control that will enable the implementation of skyrmion diffusion-based applications such as the proposed non-conventional logic [21].

**References**

[1] A. Fert, V. Cros, and J. Sampaio, Nat. Nanotechnol. **8**, 152–156 (2013).
[2] X. Zhang, M. Ezawa and Y. Zhou, Sci. Rep. **5**, 9400 (2015).
[3] Y. Huang, W. Kang, X. Zhang, Y. Zhou and Weisheng Zhao, Nanotechnology **28**, 08LT02 (2017).
[4] D. Pinna, F. Abreu Araujo, J.-V. Kim, V. Cros, D. Querlioz, P. Bessiere, J. Droulez, and J. Grollier, Phys. Rev. Appl. **9**, 064018 (2018).
[5] A. Fert, N. Reyren & V. Cros, Nat. Rev. Mater. **2**, 17031 (2017).
[6] J. Zázvorka, F. Jakobs, D. Heinze, N. Keil, S. Kromin, S. Jaiswal, K. Litzius, G. Jakob, P. Virnau, D. Pinna, K. Everschor-Sitte, L. Rózsa, A. Donges, U. Nowak and M. Kläui, Nat. Nanotechnol. **14**, 658–661 (2019).
[7] I. Dzyaloshinsky, J. Phys. Chem. Solids **4**, 241–255 (1958).
[8] T. T. Moriya, Phys. Rev. **120**, 91–98 (1960).
[9] U. K. Rößler, A. N. Bogdanov and C. Pfleiderer, Nature **442**, 797–801 (2006).
[10] S. Mühlbauer, B. Binz, F. Jonietz, C. Pfleiderer, A. Rosch, A. Neubauer, R. Georgii, P. Böni, Science **323**, 915–919 (2009).
[11] X. Z. Yu, N. Kanazawa, Y. Onose, K. Kimoto, W. Z. Zhang, S. Ishiwata, Y. Matsui and Y. Tokura, Nat. Mater. **10**, 106–109 (2011).
[12] K. Everschor-Sitte, J. Masell, R. M. Reeve and M. Kläui, J. Appl. Phys. 124, 240901 (2018).
[13] W. Jiang, G. Chen, K. Liu, J. Zang, S. G. E. te Velthuis and A. Hoffmann, Phys. Rep. **704**, 1–49 (2017).
[14] R. Wiesendanger, Nat. Rev. Mater. **1**, 16044 (2016).
[15] S. Finizio, K. Zeissler, S. Wintz, S. Mayr, T. Weßels, A. J. Huxtable, G. Burnell, C. H. Marrows and J. Raabe, Nano Lett. **19**, 10, 7246-7255 (2019).
[16] W. Jiang, Z. Xichao, G. Yu, W. Zhang, X. Wang, M. B. Jungfleisch, J. E. Pearson, X. Cheng, O. Heinonen, K. L. Wang, Y. Zhou, A. Hoffmann and S.G.E. te Velthuis, Nat. Phys. **13**, 162-169 (2017).
[17] S. Woo, K. Litzius, B. Krüger, M.-Y. Im, L. Caretta, K. Richter, M. Mann, A. Krone, R. M. Reeve, M. Weigand, P. Agrawal, I. Lemesh, M.-A. Mawass, P. Fischer, M. Kläui and G. S. D. Beach, Nat. Mater. **15**, 501–506 (2016).
[18] K. Litzius, I. Lemesh, B. Krüger, P. Bassirian, L. Caretta, K. Richter, F. Büttner, K. Sato, O. A. Tretiakov, J. Förster, R. M. Reeve, M. Weigand, I. Bykova, H. Stoll, G. Schütz, G. S. D. Beach and M. Kläui, Nat. Phys. **13**, 170–175 (2017).
[19] K. Litzius, J. Leliaert, P. Bassirian, D. Rodrigues, S. Kromin, I. Lemesh, J. Zazvorka, K.-J. Lee, J. Mulkers, N. Kerber, D. Heinze, N. Keil, R. M. Reeve, M. Weigand, B. V. Waeyenberge, G. Schütz, K. Everschor-Sitte, G. S. D. Beach and M. Kläui, Nat. Electronics. **3**, 30–36 (2020).
[20] C. Schütte, J. Iwasaki, A. Rosch and N. Nagaosa, Phys. Rev. B **90**, 174434 (2014).
[21] T. Nozaki, Y. Jibiki, M. Goto, E. Tamura, T. Nozaki, H. Kubota, A. Fukushima, S. Yuasa and Y. Suzuki, Appl. Phys. Lett. **114**, 012402 (2019).
[22] Y. Jibiki, M. Goto, E. Tamura, J. Cho, H. Nomura, T. Srivastava, W. Lim, S. Auffret, C. Baraduc, H. Bea and Y. Suzuki, arXiv:1909.10130 (2019).
[23] L. Zhao, Z. Wang, X. Zhang, J. Xia, K. Wu, H.-A. Zhou, Y. Dong, G. Yu, K. L. Wang, X. Liu, Y. Zhou and W. Jiang, arXiv:1901.08206 (2019).
[24] F. Peper, J. Lee, J. Carmona, J. Cortadella and K. Morita, ACM J. Emerg. Technol. Comput. Syst. **9**, 3 (2013).
[25] E. Nelson, Dynamical Theories of Brownian Motion (Princeton Univ. Press, Princeton, NJ, 1972).
[26] A. Einstein, Ann. Phys. **17**, 549 (1905).
[27] J. Happel and H. Brenner, Low Reynolds Number Hydrodynamics (Kluwer, Dordrecht, Netherlands, 1991).

[28] Y. Han, A. M. Alsayed, M. Nobili, J. Zhang, T. C. Lubensky and A. G. Yodh, Science **314**, 626-30 (2006).

[29] I. Lemesh, K. Litzius, M. Böttcher, P. Bassirian, N. Kerber, D. Heinze, J. Zázvorka, F. Büttner, L. Caretta, M. Mann, M. Weigand, S. Finizio, J. Raabe, M.-Y. Im, Hermann Stoll, G. Schütz, B. Dupé, M. Kläui and G. S. D. Beach, Adv. Mater. **30**, 1805461 (2018).

[30] J.-Y. Tinevez, N. Perry, J. Schindelin, G. M. Hoopes, G. D. Reynolds, E. Laplantine, S Y. Bednarek, S. L. Shorte and K. W. Eliceiri, Methods **115**, 80–90 (2017).

[31] A. A. Thiele, Phys. Rev. Lett. **30**, 230 (1973).

[32] J. Miltat, S. Rohart, and A. Thiaville, Phys. Rev. B **97**, 214426 (2018).

[33] M. Weißenhofer and U. Nowak, Phys. Rev. B **99**, 224430 (2019).

[34] J. Xia, X. Zhang, M. Ezawa, Q. Shao, X. Liu and Y. Zhou, Appl. Phys. Lett. **116**, 022407 (2020).

[35] T. Kamppeter, F. G. Mertens, E. Moro, A. Sánchez and A. R. Bishop, Phys. Rev. B **59**, 11349 (1999).

[36] S.-Z. Lin and A. Saxena, Phys. Rev. B **92**, 180401(R) (2015).

[37] A. Vansteenkiste, J. Leliaert, M. Dvornik, M. Helsen, F. Garcia-Sanchez and B. V. Waeyenberge, AIP Advances **4**, 107133 (2014).

[38] J. Jena, B. Göbel, T. Ma, V. Kumar, R. Saha, I. Mertig, C. Felser and S. S. P. Parkin, Nat. Commun. **11**, 1115 (2020).

**Acknowledgments**

N.K., K.L. and M.K. gratefully acknowledge financial support by the Graduate School of Excellence Materials Science in Mainz (MAINZ, GSC266). The Deutsche Forschungsgemeinschaft (DFG, German Research Foundation) has funded this project in Mainz and Konstanz through the priority programme SPP 2137 on Skyrmionics and M.K. and the group in Mainz through projects 290319996/TRR173 and 290396061/TRR173.

**Author Contributions**

M.K. and U.N. proposed and supervised the study. J.Z. and K.L. fabricated the multilayer samples. N.K., K.R and J.Z. prepared the measurement set-up and conducted the experiments using the Kerr microscope. N.K. and K.R. evaluated the experimental data with the help of K.L. and J.Z. M.W. performed the analytical calculations and numerical micromagnetic simulations. N.K. drafted the manuscript with the help of M.W., M.K. and U.N. All the authors commented on the manuscript.

**Competing interests**

The authors declare no competing interests.